\documentclass[%
reprint,
superscriptaddress,
showpacs,
amsmath,amssymb,
aps,
pra,
]{revtex4-1}

\usepackage{graphicx}
\graphicspath{{/}}
\DeclareGraphicsExtensions{.pdf,.png,.jpg}
\usepackage{dcolumn}% Align table columns on decimal point
\usepackage{bm}% bold math
\usepackage{hyperref}% add hypertext capabilities
\usepackage{amsmath,amssymb,amsthm}
\DeclareMathOperator{\Tr}{Tr}

\begin{document}

%\preprint{APS/123-QED}

\title{Security of subcarrier wave quantum key distribution against the collective beam-splitting attack}%

\author{A. V. Kozubov}   \email{avkozubov@corp.ifmo.ru}
\affiliation{Department of Photonics and Optical Information Technology, ITMO University, Saint Petersburg, Russia}%

\author{A. A. Gaidash}%
\affiliation{Department of Photonics and Optical Information Technology, ITMO University, Saint Petersburg, Russia}%

\author{A. V. Gleim}%
\affiliation{Department of Photonics and Optical Information Technology, ITMO University, Saint Petersburg, Russia}%

\author{G. P. Miroshnichenko}%
\affiliation{Department of Photonics and Optical Information Technology, ITMO University, Saint Petersburg, Russia}%

\author{D. B. Horoshko}%
\affiliation{Univ. Lille, CNRS, UMR 8523, Physique des Lasers Atomes et Molecules (PhLAM), F-59000 Lille, France}
\affiliation{B. I. Stepanov Institute of Physics, NASB, Nezavisimosti Avenue 68, Minsk 220072, Belarus}%

\date{\today}

\begin{abstract}
We consider a subcarrier wave quantum key distribution (QKD) system, where the quantum encoding is carried by weak sidebands generated to a coherent optical beam by means of an electrooptic phase modulation. We study the security of two protocols, B92 and BB84, against one of the most powerful attacks on the systems of this class: the collective beam splitting attack. We  show that a subcarrier wave QKD system with realistic parameters is capable of distributing a cryptographic key over large distances. We show also that a modification of the BB84 protocol, with discrimination of only one state in each basis, performs not worse than the original BB84 protocol for this class of QKD systems, which brings a significant simplification to the development of cryprographic networks on the basis of considered technique.  
\end{abstract}

\maketitle

%\tableofcontents

\section{\label{sec:intro}Introduction}
Growing interest to the quantum key distribution (QKD) systems \cite{Gisin02,Scarani09,Alleaume14} in the last decades has led to emergence of a large number of experimental works dedicated to the development of reliable QKD setups suitable for everyday operation in existing telecommunication networks. Among them stand subcarrier wave (SCW) QKD systems, the most valuable feature of which is exceptionally efficient use of the quantum channel bandwidth and capability of signal multiplexing by adding independent sets of quantum subcarriers to the same carrier wave. It makes SCW QKD systems perfect candidates as backbone of multiuser quantum networks.

In the SCW QKD system a strong monochromatic wave, produced by a laser, is modulated to produce weak sidebands whose phase with respect to the strong coherent wave encodes the quantum information. Various protocols can be realized with this technique, the most popular ones being the BB84 protocol \cite{Bennett92jc}, using four phase values, and the B92 protocol \cite{Bennett92}, using just two phases. A realization of the B92 protocol with phase modulation has been demonstrated by Merolla group \cite{Merolla99a,Merolla99b}. A realization of the BB84 protocol has been demonstrated by the same group with the help of amplitude rather than phase modulators \cite{Merolla02}; the replacement of the phase modulation by a more technically complicated amplitude modulation being necessary for decoding all the states of the protocol at the side of receiver. The latter approach, combined with the employment of several microwave frequencies in the amplitude modulators by the technique of subcarrier multiplexing resulted recently in a significant increase of the key generation rate \cite{Mora12}. It was shown \cite{Guerreau05} that monitoring the intensity of the strong wave can provide additional security to the protocol, being a realization of the method of ``strong reference'' \cite{Bennett92}, suggested in the early days of quantum cryptography for fighting the most dangerous attacks, and proven recently to provide unconditional security for QKD with weak coherent states \cite{Koashi04}. A variant of the BB84 protocol with phase modulation, allowing the receiver to decode only one of the two states in each basis, has been recently realized by some of us \cite{Gleim16}.

Notwithstanding the large experimental effort for building SCW QKD systems, the analysis of their security still requires special consideration. In this article we explore the security of the B92 and BB84 protocols against one but very powerful attack: the collective beam-splitting (CBS) attack. We analyse this attack and calculate the secure key generation rate for given protocols in its presence. The CBS attack is not limited by the employment of the strong reference, thus the obtained result is quite general and remains valid even for strong-reference-enhanced versions of the protocols. We show also that the secure key generation rate for the BB84 protocol with one state decoding (BB84-OSD) is the same as for BB84 with both states decoded, which allows the developers of the QKD networks to employ a relatively simple phase modulation in the future.

Our calculation of the secure key rate is based on the well-known Devetak-Winter bound \cite{Devetak05,Scarani09}, applicable to the case of collective attacks in one-way protocols of QKD with independent identically distributed information carriers, to which the considered protocols of SCW QKD belong. We employ the recently developed quantum model of electro-optical phase modulation \cite{Miroshnichenko17}, which allows us to deduce the states of sidebands in the quantum channel after the modulation, and their states after the demodulation before the detection. This model has an advantage of being applicable in the case of relatively high modulation index, where tens of sidebands contain non-negligible amount of photons.

This paper is organized as follows. Section II gives a description of the protocols implemented in SCW QKD device and builds the model of the quantum channel. In Section III we calculate the quantum bit error rate as function of loss in the quantum channel, which take into consideration the quantum efficiency and the dark count rate of the photodetector.  In Section IV we consider the attacks on the protocols and in Section V we find the secure key rate dependence on the channel length for different sets of SCW QKD parameters. Section VI concludes the article.

\section{\label{sec2}Operation principles of SCW QKD}

\subsection{The setup and the protocols}
\begin{figure}
\includegraphics[width=\columnwidth]{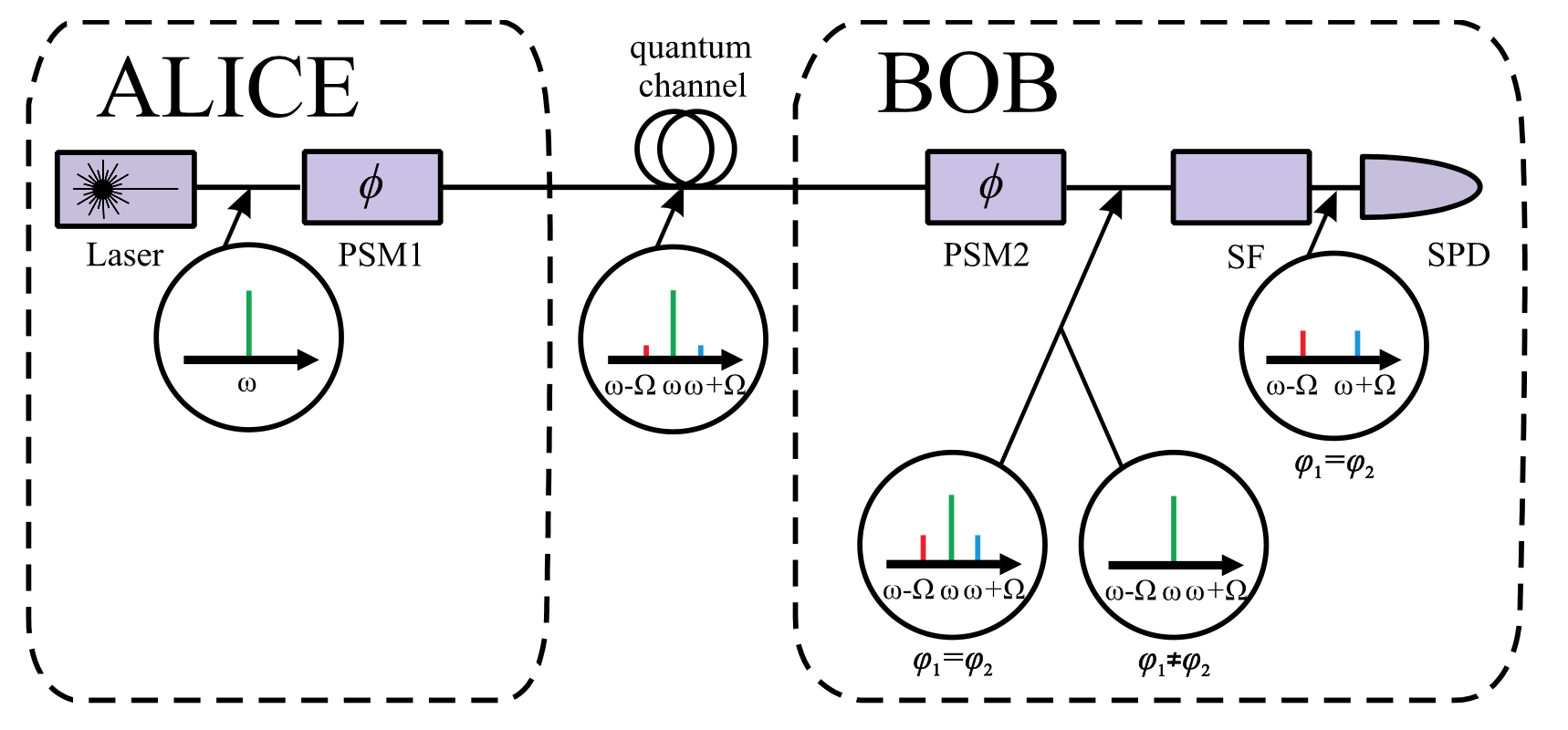}
\caption{Schematic presentation of the SCW QKD setup. PSM is a phase modulator; SF is a spectral filter, removing the central frequency; SPD is the single photon detector. The diagrams in circles show the spectrum in the corresponding part of the setup. Only two sidebands are shown in the spectra for simplicity.}
\label{fig1}
\end{figure}

We consider two protocols of SCW QKD sharing the same experimental setup, which is schematically shown in Fig.~\ref{fig1}. The laser source produces a coherent monochromatic light beam with the optical frequency $\omega$, serving as the carrier wave of the setup. The sender Alice modulates this beam by means of a travelling-wave phase modulator, with the frequency of the microwave field $\Omega$ and its phase $\varphi_A$. As a result of phase modulation, the field at the output of the modulator acquires sidebands at frequencies $\omega_k=\omega+k\Omega$, where we limit ourselves to $2S$ sidebands and let the integer $k$ to run in the limits $-S\le k\le S$. The modulation index and the intensity of the carrier wave are chosen so that the total number of photons in the sidebands is less than unity, thus providing non-orthogonality of the used set of states, required by the no-cloning theorem, lying in the heart of the QKD security. The phase $\varphi_A$ is constant in a transmission window of duration $T$, but changes randomly within a predefined set in the next window. The value of this phase is written on the relative phase between the sidebands and the carrier wave, and thus encodes the bit, sent by Alice. In this article we consider two protocols, differing by the set of phases, used by Alice. The B92 \cite{Bennett92} protocol employs only two non-orthogonal states and $\varphi_A\in\{0,\pi\}$, encoding the logical 0 and 1 respectively. The BB84 protocol \cite{Bennett92jc} uses four states, split in two bases: $\varphi_A\in\{0,\pi\}$ corresponds to the basis 0, while $\varphi_A\in\{\pi/2,3\pi/2\}$ corresponds to the basis 1, the first state of each basis encoding the logical 0, and the second one encoding the logical 1. In both protocols the carrier wave propagates together with the sidebands and can serve as ``strong reference'', which was suggested for B92 protocol by its author Bennett \cite{Bennett92}, and can be extended to the BB84 protocol by analogy. Monitoring the power of the strong reference helps to fight the attacks employing measurement of the quantum carriers in the quantum channel and suppression of them in the case of unfavorable outcome, like the photon number splitting (PNS) attack and the unambiguous state discrimination (USD) attack. However, in the present article we do not consider the additional enhancement of security provided by the strong reference method, leaving it to a separate study.

The encoded states together with the carrier wave are sent to the receiver Bob, who applies a similar phase modulation to the received beam with the microwave phase $\varphi_B$ in each transmission window, and then directs all the sidebands to a single-photon detector (SPD). The set of phases used by Bob is the same as that of Alice in both protocols. The decoding is based on the fact, that each time Alice and Bob use different phases from the same set $\{0,\pi\}$ or $\{\pi/2,3\pi/2\}$, so that $\varphi_A-\varphi_B=\pm\pi$, the sidebands after the Bob's modulator are in the vacuum state and the SPD produces no click, except for dark counts. In the B92 protocol Bob decodes a bit value in the transmission windows where his SPD clicks, and this value corresponds to his phase $\varphi_B$. In the BB84-OSD protocol Bob waits while Alice announces the bases used for each bit by a public channel, and then decodes the bit value in the transmission windows where he used the same basis and where his SPD clicked, this value again corresponding to his phase $\varphi_B$. In this protocol Bob decodes only one state of the basis. For example, if Alice uses $\varphi_A=0$, Bob decodes this bit only if he uses the phase $\varphi_B=0$. The phase $\varphi_B=\pi$, however belonging to the same basis, does not result (in the ideal case) in a click of Bob's detector. We show in Sec. V that this modification of the BB84 protocol does not affect the secure key generation rate.

\subsection{The secure key generation rate}

The described above protocols belong to the class of one-way protocols of QKD with independent identically distributed information carriers \cite{Scarani09}. The secure key generation rate $K$ for the protocols of this class in the presence of collective attacks is lower bounded by the Devetak-Winter bound \cite{Scarani09,Devetak05}:
\begin{equation}\label{K}
K=\nu_S P_B \left[1-\mathrm{leak}_{EC}(Q)-\max_E\chi(A:E)\right],
\end{equation}
where $\nu_S$ is the repetition rate, in our case $\nu_S=T^{-1}$; $P_B$ is the probability of successful decoding and accepting a bit in one transmission window; $Q$ is the quantum bit error rate (QBER), the probability that a bit, accepted by Bob is erroneous; $\mathrm{leak}_{EC}(Q)$ is the amount of information revealed by Alice by the public channel for the sake of the error correction, which depends on QBER and is limited by the Shannon bound: $\mathrm{leak}_{EC}(Q)\ge h(Q)$, where $h(Q)=-Q\log_2Q-(1-Q)\log_2(1-Q)$ is the binary Shannon entropy.

The quantity $\chi(A:E)$ in Eq.~(\ref{K}) is the Holevo information \cite{Holevo82}, giving the upper bound for the information accessible to the eavesdropper Eve in a given collective attack. In this class of attacks Eve realizes interaction of her ancilla with each information carrier in the quantum channel (light in one transmission window in our case), stores the ancillas for the entire transmitted block in a quantum memory, and waits while Alice and Bob finish the post-processing of their key. Afterwards Eve measures collectively all the ancillas of the block, taking into account all the information collected from the public channel. The best measurement cannot give her more information (per bit) than
\begin{equation}\label{chi}
\chi(A:E) = S(\rho) - \sum_k p_k S(\rho_k),
\end{equation}
where the index $k$ enumerates the possible states in the quantum channel, $\rho_k$ is the state of the ancilla under condition that $k$th state was attacked, $p_k$ is the weight of the $k$th state, $\rho = \sum_k p_k \rho_k$ is the unconditional state of ancilla, and $S(\rho) = -\Tr\{\rho\log_2\rho\}$ is the von Neumann entropy. The accessible information in Eq.~(\ref{K}) is maximized over all possible attacks by Eve, which is almost impossible to realize by considering various attacks one by one. A different approach to finding the secure key rate is connected to considering an equivalent protocol of entanglement distillation \cite{Shor00}, and proving thus the unconditional security of the given protocol. Unfortunately, such an approach has not been yet applied to the SCW QKD without strong reference. In Sec. IV we calculate the quantity, Eq.~(\ref{chi}), for just one, but very powerful attack, the CBS attack.

\subsection{The quantum channel}

The information channel between Alice and Bob, including quantum encoding, transmission via quantum channel, and quantum decoding, for each choice of basis has two input values, Alice's bit $x=0,1$, and three output values, Bob's bit $y=0,1$ and the inconclusive result $y=2$ where Bob's detector does not click. The absence of click is caused by the vacuum component in the state of the sidebands, and also by the possibility of destructive interference in the case Bob guesses the basis but not the state in BB84-OSD. The channel is completely determined by the matrix $P(y|x)$, the conditional probability of the Bob's outcome $y$ when Alice sends $x$. We accept that Alice's bit is random and its two values are equiprobable, which is known to maximize the channel capacity.

\begin{figure}[h]
\includegraphics[width=\columnwidth,keepaspectratio]{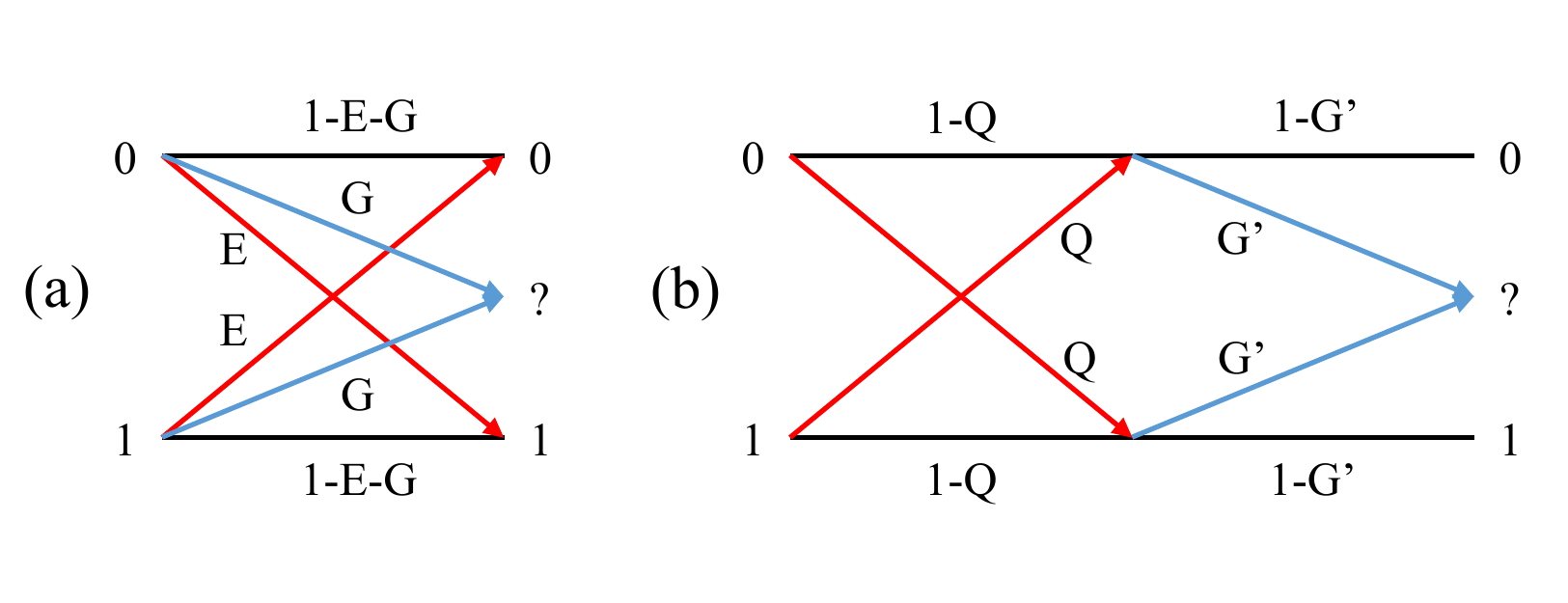}
\caption{Diagrams of (a) the binary symmetric error and erasure channel and (b) the equivalent cascade of a symmetric binary channel and an erasure channel. The question mark denotes the inconclusive result. }
\label{fig2:channel}
\end{figure}

In the case where the probabilities of error and loss are independent of the input values, such a channel represents a binary symmetric error and erasure (BSEE) channel \cite{Cover91}. For such a channel we can write $E=P(0|1)=P(1|0)$, $G=P(2|0)=P(2|1)$, and these two parameters determine completely the channel. The diagram of this channel is shown in Fig.~\ref{fig2:channel}a and its capacity is given by
\begin{eqnarray}\label{Cap}
C&=&1-G-(1-G)\log_2(1-G)+E\log_2(E)+\nonumber\\
&+&(1-G-E)\log_2(1-G-E).
\end{eqnarray}

It should be noted, that the ``error probability'' $E$ of BSEE channel is not the QBER value $Q$, entering Eq.~(\ref{K}), because the latter is the probability of error under condition of conclusive measurement outcome. To find the value of QBER, we represent BSEE channel as two cascaded channels \cite{Cover91}: a symmetric binary channel with error probability $Q$ and an erasure channel with the erasure probability $G'$, see Fig.~\ref{fig2:channel}b. It is easy to find that the two representations are equivalent, i.e, have the same matrix $P(y|x)$, if $G'=G$ and $Q(1-G)=E$. It is easy to verify also that the capacity of the channel, given by Eq.~(\ref{Cap}), can be rewritten as $C=C_1C_2$, where $C_1=1-h(Q)$ is the capacity of the binary symmetric channel, and $C_2=1-G$ is the capacity of the erasure channel. In the next section we calculate the values of $E$ and $G$, and as consequence, the QBER, from the explicit expressions for the quantum states, used in the SCW QKD system.

\section{\label{sec:3}Quantum bit error rate}

The states of the multimode optical field at the entrance to the quantum channel can be found by the quantum model of electro-optical phase modulator developed in Ref.~\cite{Miroshnichenko17}. The model takes into consideration $2S+1$ modes of the optical field with frequencies $\omega+k\Omega$, where the integer $k$ varies as $-S\le k\le S$. The input state of the Alice's modulator is $|\sqrt{\mu_0}\rangle_0\otimes|\mathrm{vac}\rangle_{SB}$, where $|\mathrm{vac}\rangle_{SB}$ is the vacuum state of the sidebands and $|\sqrt{\mu_0}\rangle_0$ is a coherent state of the carrier wave with the amplitude $\sqrt{\mu_0}$, determined by the average number of photons in a transmission window: $\mu_0=PT/(\hbar\omega)$, $P$ being the power of the laser beam. The phase of the coherent state of the carrier wave is accepted to be zero and all other phases are calculated with respect to this phase. The state of the field at the output of the modulator is a multimode coherent state
\begin{equation}
|\psi_0(\varphi_A)\rangle = \bigotimes_{k=-S}^S|{\alpha_k(\varphi_A)}\rangle_k,
\end{equation}
with the coherent amplitudes
\begin{equation}\label{alpha}
\alpha_k(\varphi_A)=\sqrt{\mu_0}d^S_{0k}(\beta)e^{-i(\theta_1+\varphi_A)k},
\end{equation}
where $\theta_1$ is a constant phase and $d^S_{nk}(\beta)$ is the Wigner d-function, appearing in the quantum theory of angular momentum \cite{Varshalovich88}. The argument of the d-function $\beta$ is determined by the modulation index $m$, and disregarding the dispersion of the modulator medium this dependence can be written as \cite{Miroshnichenko17}:
\begin{equation}
\beta =\frac{2m}{2S+1}.
\end{equation}
A remarkable property of the d-function is its asymptotic form \cite{Varshalovich88}
\begin{equation}\label{asymp}
d^S_{nk}(m/S)\underset{S\to\infty}\longrightarrow J_{n-k}(m),
\end{equation}
where $J_{n}(x)$ is the Bessel function of the first kind. This asymptotic form corresponds to a conventional description of the phase modulation with an infinite number of sidebands, leading in the quantum case to unphysical results because of appearance of negative frequencies, see discussion in Ref.~\cite{Capmany10}.

After passing the distance $L$ in the quantum channel (optical fiber), the states of all spectral components are attenuated. The transmission coefficient of the quantum channel is $\eta(L)=10^{-\xi L/10}$, where $\xi$ is the fibre loss per unit length. The state of the optical field in one transmission window at the entrance to the Bob's module is
\begin{equation}\label{psiL}
|\psi_L(\varphi_A) \rangle=\bigotimes_{k=-S}^S|\sqrt{\eta(L)}\alpha_k(\varphi_A)\rangle_k.
\end{equation}

The microwave field in the Bob's phase modulator has the same frequency $\Omega$ as that of the Alice's one, but a different phase $\varphi$. Additional field is produced in this modulator on the same sideband frequencies $\omega+k\Omega$, which interferes with the field already present on these frequencies. The resulting state of the field is a multimode coherent state \cite{Miroshnichenko17}
\begin{equation}\label{psiB}
|\psi_B(\varphi_A,\varphi)\rangle = \bigotimes_{k=-S}^S|{\alpha_k'(\varphi_A,\varphi)}\rangle_k,
\end{equation}
with the coherent amplitudes
\begin{equation}\label{alphaprime}
\alpha_k'(\varphi_A,\varphi)=\sqrt{\mu_0\eta(L)}d^S_{0k}(\beta')e^{-i(\theta_2+\varphi_A+\varphi)k},
\end{equation}
where the new argument of the d-function is determined by the relation
\begin{equation}\label{betaprime}
\cos{\beta'}=\cos^2{\beta}-\sin^2{\beta}\cos\left(\varphi_A-\varphi+\varphi_0\right),
\end{equation}
while $\theta_2$ and $\varphi_0$ are some constant phases determined by the construction of the phase modulator \cite{Miroshnichenko17}. Equation~(\ref{betaprime}) shows that to achieve constructive interference on the sidebands, Bob should use $\varphi_0$ as offset for his phase, and apply in his modulator the microwave phase $\varphi=\varphi_0+\varphi_B$. Then, for $\varphi_A-\varphi_B=0$, the argument of the d-function doubles: $\beta'=2\beta$, while for $\varphi_A-\varphi_B=\pm\pi$ it vanishes: $\beta'=0$. Since $d^S_{0k}(0)=\delta_{0k}$, a zero argument corresponds to the presence of photons only on the carrier frequency, all the sidebands being in the vacuum state.

The optical losses in the Bob's module can be described by the transmittance coefficient $\eta_B$. These losses can be taken into account by replacing the amplitudes determined by Eq.~(\ref{alphaprime}) with the following ones:
\begin{equation}\label{alphabar}
\bar\alpha_k(\varphi_A,\varphi_B)=\sqrt{\mu_0\eta(L)\eta_B}d^S_{0k}(\beta')e^{-i(\theta_3+\varphi_A+\varphi_B)k},
\end{equation}
where $\theta_3=\theta_2+\varphi_0$. It is unimportant if some of the optical losses took place before the phase modulation, during or after it, as soon as they are the same for all spectral components.

The spectral filtering in the Bob's module aims at removing the relatively strong carrier wave. Unfortunately, in a practical QKD system this wave can be only attenuated by the factor $\vartheta\ll1$, resulting in a replacement $\bar\alpha_0(\varphi_A,\varphi_B) \to \sqrt{\vartheta}\bar\alpha_0(\varphi_A,\varphi_B)$.

Thus, the average number of photons, arriving at the Bob's detector in the transmission window $T$ is given by the average total number of photons at all spectral components
\begin{eqnarray}\label{nph}\nonumber
n_{ph}(\varphi_A, \varphi_B)&=& \vartheta|\bar\alpha_0(\varphi_A,\varphi_B)|^2+\sum_{k\ne 0} |\bar\alpha_k(\varphi_A,\varphi_B)|^2 \\
&=&\mu_0\eta(L)\eta_B\left(1-(1-\vartheta)| d^S_{00}(\beta')|^2\right),
\end{eqnarray}
where we have used the property of the d-functions \cite{Varshalovich88}
\begin{eqnarray}\label{d-func}
\sum_{k=-S}^S d^S_{nk}(\beta)\left(d^S_{lk}(\beta)\right)^* = \delta_{nl},
\end{eqnarray}
meaning that $d^S_{nk}(\beta)$ is a unitary matrix with respect to its lower indices.

For the values $n_{ph}\ll 1$, typical for a long-distance QKD line, the probability for the SPD to produce a click in the window $T$ is \cite{Mandel95}
\begin{equation}\label{Pph}
P_{ph}(\varphi_A,\varphi_B)=\left(\eta_D \frac{n_{ph}(\varphi_A,\varphi_B)}{T} + \gamma_{dark}\right)\Delta t,
\end{equation}
where $\eta_D$ is the detector quantum efficiency, $\gamma_{dark}$ is the dark count rate, and $\Delta t = T$ for the continuous operation of the detector, but if a gating time shorter than $T$ is used, then $\Delta t$ is equal to the gating time of the detector.

Now we can calculate the parameters of the BSEE channel between Alice and Bob. These parameters are the same for both protocols B92 and BB84-OSD, and are given by the following relations
\begin{eqnarray}\label{Eval}
P_{det}(0,\pi+\Delta\varphi)&=&E,\\\label{Gval}
P_{det}(0,\Delta\varphi))&=&1-E-G,
\end{eqnarray}
where the phase $\Delta\varphi$ describes slight phase instability, caused, for instance, by jitter or phase mismatch due to non-perfect synchronization.

The QBER $Q=E/(1-G)$ can be calculated from Eqs.~(\ref{nph},\ref{Pph},\ref{Eval},\ref{Gval}) as function of the distance $L$. The calculation can be simplified by considering a sufficiently high number of sidebands $S\gg10$ and taking the limit, given by Eq.~(\ref{asymp}).

For calculations we use the experimental parameters from one of the regimes realized in Ref.~\cite{Gleim16}: $T=10$ ns, $\mu_0=4$, $m=0.319$, $\nu_S=100$ MHz, $ \Delta\varphi=5^{\circ}$, $10\lg\eta_B=6.4$ dB, $\vartheta=10^{-3}$. Two different detectors are considered: a superconducting nanowire single-photon detector (SNSPD) with $\eta_D=0.2$, $\gamma_{dark}= 20$ Hz, operating in the continuous regime, and an avalanche photodiode (APD): $\eta_D=0.125$, $\gamma_{dark}= 400$ Hz, operating in the gated regime with the gating time $\Delta t=4$ ns.

\begin{figure}[h]
\includegraphics[width=\columnwidth,keepaspectratio]{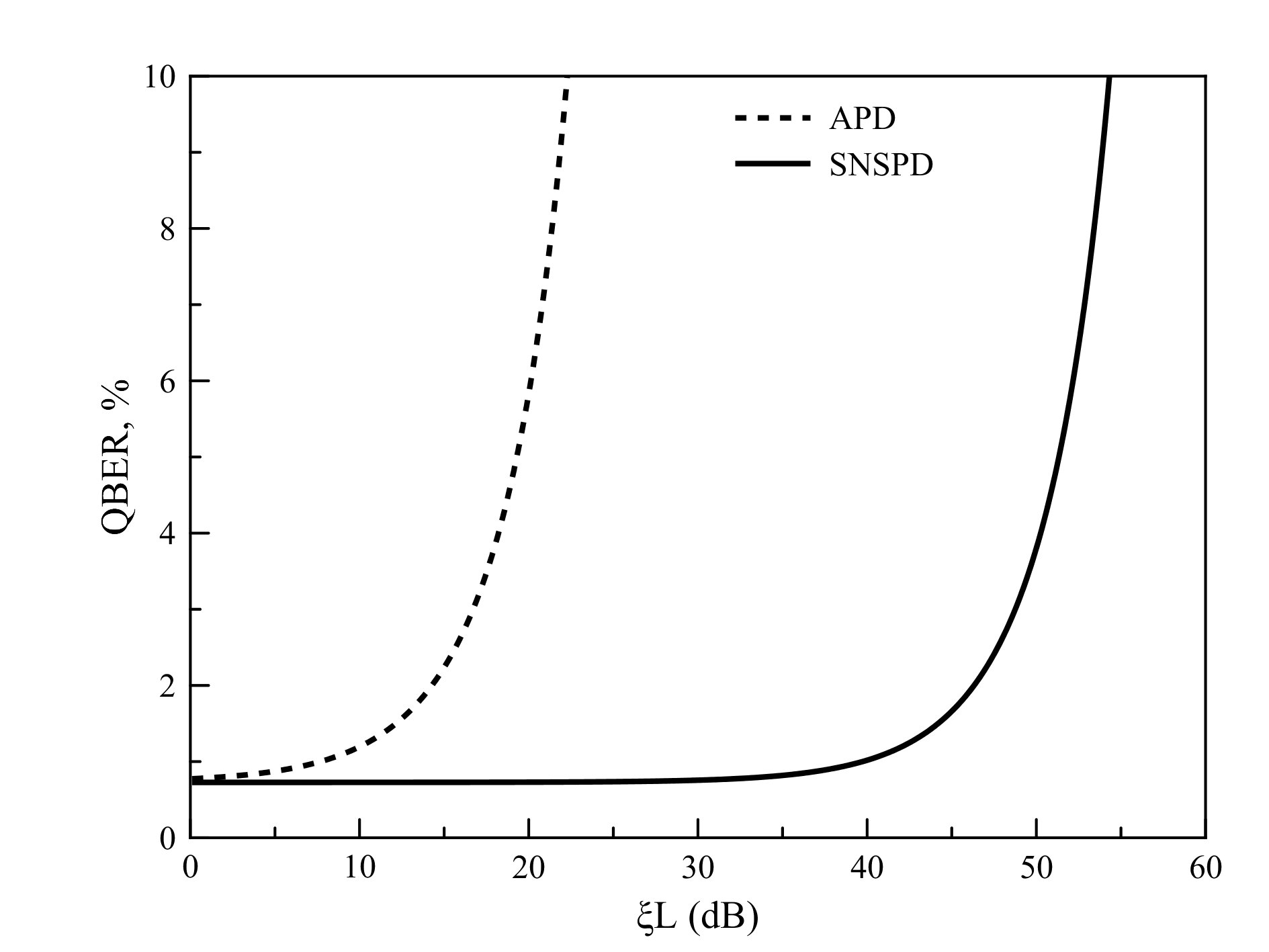}
\caption{QBER dependence on the channel loss in SCW QKD system}
\label{fig3:QBER}
\end{figure}

In Fig.\ref{fig3:QBER} we show the dependence of QBER on the optical loss $\xi L$ for two considered detectors.

For relatively low loss, while the counting rate well surpasses the dark count rate, the QBER is mainly determined by the phase instability and imperfect filtering of the carrier wave. The loss at which the counting rate becomes comparable to $\gamma_{dark}$ is the maximal loss for a given QKD system, because above this value the QBER increases rapidly, reaching values, not suitable for the error correction.

\section{\label{sec:level2}Analysis of the CBS attack}

\subsection{The possible attacks}

After Alice and Bob have successfully generated a block of shared bits of length $N$, containing some errors (the raw key), they perform error correction by disclosing $N\cdot\mathrm{leak}_{EC}(Q)$ bits, this number depending on their error-correcting protocol, but lower limited by $N\cdot h(Q)$. After having corrected all the errors they do privacy amplification by shortening their block by means of a hash function, with the aim to eliminate almost totally the potential knowledge of Eve on the shorter block (the final key). The amount to which the block should be shortened is determined by Eq.~(\ref{K}), where the second term in the square brackets corresponds to the information disclosed during the error correction stage, while the third term in the squire brackets corresponds to the upper estimate of potential information of Eve on the key.

To obtain a good upper estimate of the Eve's information, one needs to consider explicitly various attacks on the QKD line. The general scenario of the attack is as follows: Eve replaces the communication line characterized by the error rate $Q_0\le Q$ and the loss $\eta$ by a perfect errorless and lossless line and employs an eavesdropping procedure on the information carriers, introducing the same amount of error and loss as before the replacement, thus hiding her intrusion from the legitimate users, who monitor the error rate and the loss in the channel. The attacks suitable for modelling can be individual or collective, depending on the number of information carriers attacked at once. The most important individual attacks are the intercept-resend (IR) attack, introducing errors, but no loss \cite{Gisin02}, and three zero-error attacks, introducing only loss but no error \cite{Scarani09}: the PNS attack, the USD attack, and the individual beam-splitting (IBS) attack. The most important collective attacks include the CBS attack, being a quantum-memory-enhanced version of the IBS attack, and the asymmetric cloning (AC) attack, consisting in entangling an ancillary system to the information carrier by means of an asymmetric cloning machine \cite{Cerf00} or an asymmetric universal entangling machine \cite{Horoshko07}. The latter attack introduces errors but no loss.

The analysis of the previous section shows that in a SCW QKD system almost no error is caused by the transmission line, so that $Q_0\approx 0$. In the ``calibrated devices'' approach to the security analysis \cite{Scarani09} we accept that the Bob's module is calibrated for errors and loss, and Eve has no access to its performance. Then the IR and AC attacks, introducing errors, can be rather easily detected by enhancement of the measured value of QBER. The zero-error PNS and USD attacks require a suppression of the signal and the carrier wave in the case of unsuccessful measurement outcome, and can be countered by monitoring the power of the carrier wave, which is the essence of the ``strong reference'' method \cite{Bennett92,Koashi04}. The CBS attack always outperforms the IBS attack, which is its particular case, and therefore the CBS attack, in no way detectable, seems to be the most important for the security analysis of the SCW QKD system, serving the point of reference for all other attacks.

\subsection{The CBS attack}

In the CBS attack Eve inserts a beam splitter with the transmission $\eta(L)$ in the very beginning of the transmission line and sends the transmitted light to Bob via a lossless line, keeping the reflected light in a quantum memory, writing each window of duration $T$ to a separate cell of memory. After the announcement of bases (in BB84-OSD) and error correction performed by the legitimate users for a block of bits, she discards (in BB84-OSD) the memory cells, corresponding to windows where the bases used by Alice and Bob do not coincide and makes a collective measurement of the rest of the cells. Below we calculate the Holevo information, Eq.~(\ref{chi}), for the state of the information carrier in the quantum channel of an SCW QKD system.

As follows from the quantum consideration of the beam splitter \cite{ScullyZubairy97}, the state of the transmitted beam in the window $T$ is given by Eq.~(\ref{psiL}) and is identical to one which should arrive at Bob's module in the absence of eavesdropping, while the state of the reflected beam in the same window is
\begin{equation}\label{psiE}
|\psi_E(\varphi_A)\rangle = \bigotimes_{k=-S}^S|\sqrt{\bar\eta(L)}\alpha_k(\varphi_A)\rangle_k,
\end{equation}
where $\bar\eta(L)=1-\eta(L)$ and the phase $\varphi_A$ is a random member of the set, corresponding to the used protocol.

In the B92 protocol for each cell Eve needs to distinguish only two states, $|\psi_E(0)\rangle$ and $|\psi_E(\pi)\rangle$. In the BB84-OSD protocol, for the cells corresponding to Alice's choice of basis $\{\pi/2,3\pi/2\}$, Eve shifts the phase of the $k$th sideband by $\pi k/2$, which is realized by a unitary rotation of the state Eq.~(\ref{psiE}) with the evolution operator
\begin{equation}\label{U}
U=\exp\left(\frac{i\pi}2\sum_{k=-S}^Ska^\dagger_ka_k\right)
\end{equation}
where $a_k$ is the photon annihilation operator for the $k$th sideband. A unitary rotation does not change the accessible information, thus, Eve needs to distinguish the same two states as in the B92 protocol.

Since the two states to be distinguished are pure, the Holevo information, Eq.~(\ref{chi}) is given by the von Neumann entropy of the mixed state
\begin{equation}\label{rho}
\rho=\frac12|\psi_E(0)\rangle\langle\psi_E(0)|+\frac12|\psi_E(\pi)\rangle\langle\psi_E(\pi)|.
\end{equation}

The von Neumann entropy of a density operator is the Shannon entropy of its eigenvalues. The eigenvalues of operator $\rho$ are
\begin{equation}
\lambda_{1,2}=\frac{1}{2}\left(1\pm |\psi(0,\pi)|\right)
\end{equation}
where the state overlap $\psi(\varphi_1,\varphi_2)$ is calculated as
\begin{eqnarray}
&&\psi(\varphi_1,\varphi_2)=\langle \psi_E(\varphi_1)|\psi_E(\varphi_2) \rangle=\\
&&=\prod_{k=-S}^S{}_k\langle \sqrt{\bar\eta(L)}\cdot \alpha_k (\varphi_1)|\sqrt{\bar\eta(L)}\cdot \alpha_k (\varphi_2)\rangle_k\nonumber
\end{eqnarray}
Using the formula for the scalar product of two coherent states
\begin{equation}
\langle\alpha|\beta\rangle=\exp\left(-\frac{1}{2}(|\alpha|^2+|\beta|^2)+\alpha^*\beta\right)\nonumber
\end{equation}
we obtain
\begin{eqnarray}
&&\psi(\varphi_1,\varphi_2)\\
&&=\exp\left[-\frac{\bar\eta(L)}{2}\sum_{k=-S}^S(|\alpha_k(\varphi_1)|^2+|\alpha_k(\varphi_2)|^2-
2\alpha^*_k(\varphi_1)\alpha_k(\varphi_2))\right]\nonumber\\
&&=\exp\left[-\mu_0\bar\eta(L)\sum_{k=-S}^S|d^S_{0k}(\beta)|^2\left(1-e^{i(\varphi_1-\varphi_2)k}\right)\right], \nonumber
\end{eqnarray}
where we have employed Eq.~(\ref{alpha}). Using the properties of d-functions, we find
\begin{equation}
\sum_{k=-S}^S|d^S_{0k}(\beta)|^2\left(1-e^{i(\varphi_1-\varphi_2)k}\right)=1-d^S_{00}(\beta_-),
\end{equation}
where the angle $\beta_-$ is determined by the relation
\begin{equation}
\cos(\beta_-)=\cos^2(\beta)+\sin^2(\beta)\cdot\cos(\varphi_1-\varphi_2).
\end{equation}
Finally,
\begin{equation}
\psi(\varphi_1,\varphi_2)=\exp\left[-\mu_0\bar\eta(L)\left(1-d^S_{00}(\beta_-)\right)\right].
\end{equation}
where for $\varphi_1-\varphi_2=\pm\pi$ we need to substitute $\beta_-=2\beta$.

Thus, for both the B92 and the BB84-OSD protocols we obtain the Holevo information
\begin{equation}\label{chi2}
\chi(A:E) = h\left(\frac12(1-\exp\left[-\mu_0 \bar\eta(L)\left(1-d^S_{00}(2\beta)\right)\right]\right).
\end{equation}

\section{\label{sec:5} The secure key rate}
\subsection{Rate dependence on the loss}

Now we have all the necessary dependencies to calculate the secure key rate, determined by Eq.~(\ref{K}). The probability $P_B$ for decoding and accepting a bit is given by $P_B=(1-G)f$, where $f$ is the fraction of data where Bob guessed correctly the basis, equal to $\frac12$ for the BB84-OSD protocol and to 1 for the B92 protocol, and $1-G$ is the probabilty of photodetection in the window $T$ and is determined by Eqs.~(\ref{Eval},\ref{Gval}). As we have seen in the previous sections, all the functions, entering the right hand side of Eq.~(\ref{K}) are the same for the BB84-OSD and the B92 protocols, except for $f$. Thus, the secure key rate for the B92 protocol is always twice that for the BB84-OSD protocol, as far as we restrict our analysis to a CBS attack. For this reason we illustrate only the case of the latter protocol.

\begin{figure}[h]
\includegraphics[width=\columnwidth,keepaspectratio]{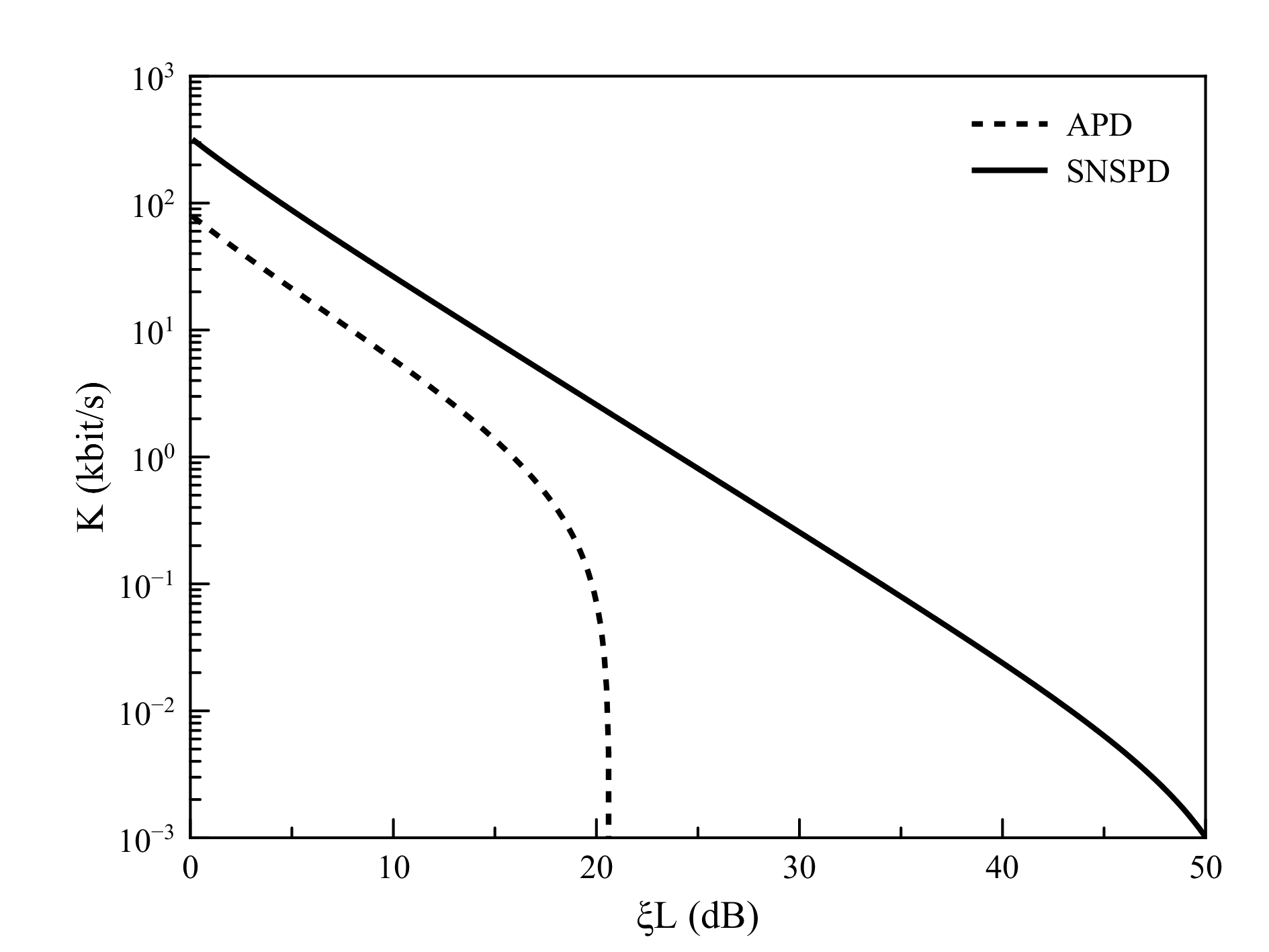}
\caption{Secure key rate dependence on channel loss in SCW QKD system}
\label{fig2:K}
\end{figure}

In Fig.~\ref{fig2:K} we show the dependence of the secure key rate on the channel loss $\xi L$ for the same setting with two detectors as in Sec.~\ref{sec:3}. The corresponding distance can be easily calculated using the value $\xi=0.18$ dB/km typical for the telecommunication fibre.

We note that for rather low repetition rate, $T^{-1}=100$ MHz, we can achieve high values of distances and secure key rates. In spite of low secure distances with APD it is still an effective solution for up to 100 km fiber lines due to its easier maintenance compared to SNSPD.

\subsection{Optimal modulation depth}

The depth of the phase modulation in the Alice's module determines the total number of photons in the sidebands of the field, entering the communication line and being the object of the Eve's attack.   
The mean photon number in all the sidebands is an important parameter, commonly used for characterizing the regime of a QKD system \cite{Gisin02,Scarani09} and it can be found as following:
\begin{equation}\label{mu}
\mu=\sum_{k\ne0}\left|\alpha_k(\varphi_A)\right|^2 = \mu_0\cdot(1-|d^S_{00}(\beta)|^2)\approx\mu_0\cdot(1-J_0(m)^2)
\end{equation}
where the last expression is the asymptotic form for a sufficiently large number of sidebands.

Higher values of $\mu$ correspond to higher counting rate of the Bob's detector and are very attractive from the practical point of view. However, the information available to Eve is also growing with $\mu$, reaching 100 \% in the limit where $\mu\gg1$ and the states of the sidebands corresponding to different values of $\varphi_A$ become almost orthogonal.

\begin{figure}[h]
\includegraphics[width=9cm,height=7cm,keepaspectratio]{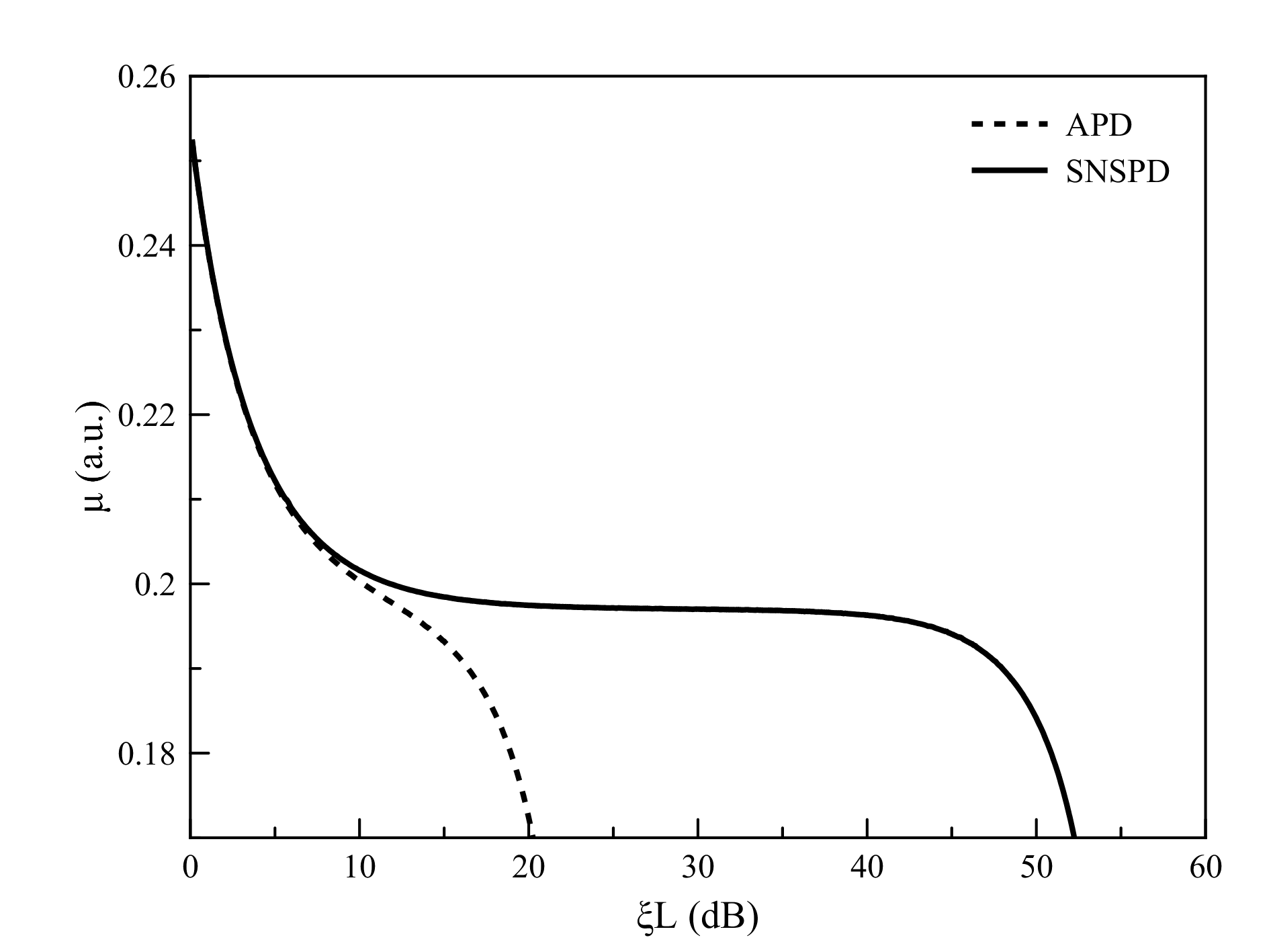}
\caption{Mean photon number dependence on losses in the channel providing the maximum value of the secret key rate}
\label{fig5:mu}
\end{figure}

The optimal value of the modulation depth $m$ and therefore $\mu$ can be found by considering the key generation rate as function $K(\mu,L)$ and finding the value $\mu(L)$ which maximizes this function for a given $L$. The numerically found dependence is presented in Fig.~\ref{fig5:mu}. According to it given value of $\mu$ was chosen.

We see that the practical value for a long-distance QKD with SNSPD is $\mu=0.2$, and higher values of the mean photon number, considered in the literature \cite{Guerreau05,Klimov17}, are not secure against the CBS attack.

\subsection{Secure key rate of BB84 versus BB84-OSD}

Let us divide all the sidebands of the field in a SCW QKD system into the upper sidebands with frequencies $\omega+k\Omega$, $k\in[1,S]$ and the lower sidebands with frequencies $\omega+k\Omega$, $k\in[-S,-1]$.

The demodulation process in the Bob's module, can be described by a demodulation operator $D(\varphi_B)$, mapping the state arriving to his module, Eq.~(\ref{psiL}), to the state given by Eq.~(\ref{psiB}). In the case $\varphi_B=0$ and if the basis is guessed correctly, this mapping is
\begin{eqnarray}\label{D0}
D(0)|\psi_L(0)\rangle &=& |\mu_s\rangle_+\otimes|\mu_s\rangle_-\otimes|\bar\mu_c\rangle_0,\\\label{Dpi}
D(0)|\psi_L(\pi)\rangle &=& |\mathrm{vac}\rangle_+\otimes|\mathrm{vac}\rangle_-\otimes|\bar\mu\rangle_0,
\end{eqnarray}
where $|\mu_s\rangle_+$ and $|\mu_s\rangle_-$ are multimode coherent states of upper and lower sidebands respectively with the same mean photon number $\mu_s$ each, defined as
\begin{eqnarray}\label{c+}
|\mu_s\rangle_+ &=& \bigotimes_{k=1}^S|{\alpha_k'(0,\varphi_0)}\rangle_k,\\\label{c-}
|\mu_s\rangle_- &=& \bigotimes_{k=-S}^{-1}|{\alpha_k'(0,\varphi_0)}\rangle_k,
\end{eqnarray}
while $|\bar\mu_c\rangle_0$ and $|\bar\mu\rangle_0$ are coherent states of the carrier wave with the mean photon numbers $\bar\mu_c=|\alpha_0'(0,\varphi_0)|^2$ and $\bar\mu=|\alpha_0'(\pi,\varphi_0)|^2$ respectively. It can be obtained from Eq.~(\ref{alphaprime}) that $\bar\mu=\bar\mu_c+2\mu_s$, which has a simple physical meaning: the demodulation process preserves the total number of photons in the field.

Let us consider a modification of the protocol BB84-OSD, where Bob has two detectors: one for the upper sidebands and one for the lower ones, and applies a demodulation operator $D'(\varphi_B)$ to the state of the field arriving to his module, Eq.~(\ref{psiL}), directing the photons either to the upper sidebands or to the lower ones depending on the phase $\varphi_B$ (under condition he uses the same basis as Alice). Here we consider only operators resulting in a linear transformation of the field, and therefore mapping coherent states onto coherent ones. In the case $\varphi_B=0$ the mapping provided by the demodulation should be as follows:
\begin{eqnarray}\label{Dprime0}
D'(0)|\psi_L(0)\rangle &=& |\mu_s'\rangle_+\otimes|\mathrm{vac}\rangle_-\otimes|\bar\mu_c'\rangle_0,\\\label{Dprimepi}
D'(0)|\psi_L(\pi)\rangle &=& |\mathrm{vac}\rangle_+\otimes|\mu_s'\rangle_-\otimes|\bar\mu_c'\rangle_0,
\end{eqnarray}
where all the states are coherent and the mean numbers of photons are determined by their arguments. In this case Bob can distinguish the cases of $\varphi_a=0$ and $\varphi_a=\pi$ by observing a click on the corresponding detector, which would correspond to a realization of the BB84 protocol. Let us calculate the corresponding counting rate.

From the unitarity of the operators $D(\varphi_B)$ and $D'(\varphi_B)$ we have
\begin{equation}\label{unit}
\langle\psi_L(\pi)| D^{\dagger}(0)D(0)|\psi_L(0)\rangle = \langle\psi_L(\pi)| D^{'\dagger}(0)D'(0)|\psi_L(0)\rangle,
\end{equation}
wherefrom, with the help of Eqs.~(\ref{D0},\ref{Dpi},\ref{Dprime0},\ref{Dprimepi}), we obtain
\begin{equation}\label{DD}
\exp\left\{-\mu_s-\frac12\left(\sqrt{\bar\mu}-\sqrt{\bar\mu_c}\right)^2\right\}=\exp\left\{-\mu_s'\right\}
\end{equation}
or
\begin{equation}\label{mumu}
\mu_s' = \mu_s+\frac12\left(\sqrt{\bar\mu}-\sqrt{\bar\mu-2\mu_s}\right)^2 \approx \mu_s+\frac{\mu_s^2}{2\bar\mu}.
\end{equation}
We see, that in the case of low modulation index, where $\mu_s/\bar\mu\ll1$, the average number of photons in the sidebands for the BB84 protocol is almost the same as for BB84-OSD, $\mu_s' \approx \mu_s$. It means that the Bob's detectors in BB84 click with the rate $\eta_B\mu_s'$ each, while in the BB84-OSD protocol (with a perfect suppression of the carrier wave) the only detector clicks with the rate $2\eta_B\mu_s$, the total count rate being the same for two protocols.

Thus, in the regime of low modulation index there is no reason to install the second detector and employ sophisticated modulation/demodulation techniques for decoding both states of the same basis. The protocol BB84-OSD performs not worse than BB84 and is technically significantly simpler. For the same reasons an increase of the number of bases or the number of phases in each basis is not expected to increase the secure key rate \cite{Horoshko02}. 

\section{\label{sec:conclusions}Conclusions}

In this work we have calculated the secure key generation rate for two protocols of SCW QKD in the presence of the CBS attack in the quantum channel and have shown that a SCW QKD system allows a secure distribution of cryptographic key over large distances. It was shown that the optimal mean photon number value in the system is $\mu \approx 0.2$. We have found that the main limiting factors for a long distance communication are the dark counts of photodetector and the fraction of photons remaining at the carrier frequency that reach the detector. It should be noted that for a more accurate QBER estimations a more advanced model of the quantum channel can be considered, for instance, including the partial loss of coherence between the sidebands propagating in the fibre.

We have shown also that the version of the BB84 protocol with detection of just one of the states in each basis has a performance not worse than that of the full BB84 protocol. Also, the key generation rate of the B92 protocol is double that of the BB84 protocol, as long as the analysis of attacks is limited to the CBS attack. It is possible, that other attacks, like USD attack, may be more successful against B92, which uses lesser number of states, than against BB84. Individual attacks on SCW QKD system will be a subject of a separate study.

The obtained results are important for constructing long-distance QKD links and multiuser quantum networks using SCW QKD instrumentation: the possibility of harnessing the ultra-high bandwidth for QKD is compatible with the existing fiber optical infrastructures.

\acknowledgments
This work was financially supported by the Government of Russian Federation, Grant 074-U01, by the Ministry of Education and Science of Russian Federation (project ¹ 14.578.21.0112, and contract ¹ 02.G25.31.0229) and by the Belarusian Republican Foundation for Fundamental Research.

\end{document}